# AI-Powered Non-Contact In-Home Gait Monitoring and Activity Recognition System Based on mm-Wave FMCW Radar and Cloud Computing

Hajar Abedi, Graduate Student Member, *IEEE*, Ahmad Ansariyan, Plinio P Morita, Member, *IEEE*, Alexander Wong, Senior Member, *IEEE*, Jennifer Boger, Member, *IEEE*, and George Shaker, Senior Member, *IEEE*

*Abstract*— In this paper, leveraging AI, cloud computing and radar technology, we create intelligent sensing that enables smarter applications to improve people's daily lives. We propose novel non-contact real-time cloud-based in-home activity recognition and gait monitoring systems. We present standalone IoT-based mm-wave radar systems coupled with deep learning algorithms as the basis of an autonomous in-home free-living physical activity recognition and gait monitoring system. We provide first-of-its-kind in-home real-life datasets. Using the mm-wave radar system, human spectrograms (time-varying micro-doppler patterns) are used to train deep Gated Recurrent Network (GRU) to identify physical activities performed by a subject in his/her living environment. An overall model accuracy of 93% was achieved to classify in-home physical activities of trained subjects in addition to 88% accuracy for a complete new subject. The proposed cloud-based system not only recognizes the type of activity and distinguishes waking periods from other in-home tasks but also records the activity level of the subject over time. Additionally, the system provides a record of washroom use frequency, sleep/sedentary/active/out-of-home durations, the subject's current state, as well as gait parameters. Therefore, our novel privacy-preserving system provides a complete record of a subject's in-home daily activity without requiring the subject to wear or carry on an extra device on his body.

## I. INTRODUCTION

FEATURES RELATED to gait are fundamental metrics of human motion and human health [1]. Human gait has been shown to be a useful and feasible clinical marker to determine the risk of functional decline - both mental and physical [2], [3]. Changes in gait parameters from a person's normal values often indicate deteriorating changes in health [4]. Technologies that could detect changes in people's gait patterns, especially older adults, could be used to support the detection, evaluation, and monitoring of parameters related to changes in mobility, cognition, and frailty [5]. Gait assessments could be leveraged as a clinical measurement as it is not limited to a specific health care discipline, also being a consistent and sensitive test [2]. Numerous studies have been conducted to identify the relative association between walking and functional decline in people, especially older adults (e.g., [6]–[9]). While gait parameters have been assessed and used as a clinical indicator for health status in various studies, there is no consensus for a standard measurement methodology for walking tests [2]. Moreover, most of the measurements are conducted during clinical visits [6]. However, variations in gait characteristics as a result of cognitive or other conditions may go undetected as the effect can be gradual and often goes unnoticed by the individual and/or during clinical visits [10]. Another issue related to assessing gait is that the unfamiliar setting of a clinic often causes people to (intentionally or unintentionally) change their gait patterns during clinical assessments. While systems such as GaitRite [11], [12] and Vicon [13] are currently available and can make precise measures of gait, such systems are expensive and difficult to operate, making them impractical for a clinic and not suitable for in-home monitoring. Therefore, there is a pressing need for affordable technology that can measure human gait parameters continuously, unobtrusively, and reliably if we are to get a better understanding of people's true gait and how their gait may change over time. A system is needed that can measure and analyze people's gait in their living environment, namely, at home, in hospitals or long-term care facilities.

A wearable device could be a possible solution for frequent gait assessments [14], [15], but using them requires people to

This work was supported by Microsoft Inc and MITACS.
H. Abedi is with the System Design Engineering Department, University of Waterloo, Waterloo, Ontario, Canada (e-mail: habedifi@ uwaterloo.ca).
A. Ansariyan is with the Electrical Engineering Department, University of Waterloo, Waterloo, Ontario, Canada (e-mail: ahmad.ansariyan@uwaterloo.ca).
P. P. Morita is with School of Public Health Sciences, University of Waterloo, Waterloo, Ontario, Canada (e-mail: plinio.morita@uwaterloo.ca).
J. Boger is with the System Design Engineering Department, University of Waterloo, Waterloo, Ontario, Canada (e-mail: jboger@ uwaterloo.ca).
A. Wong is with the System Design Engineering Department, University of Waterloo, Waterloo, Ontario, Canada (e-mail: alexander.wong@ uwaterloo.ca).
G. Shaker is with the Electrical Engineering Department, University of Waterloo, Waterloo, Ontario, Canada (e-mail: gshaker@ uwaterloo.ca).

want and remember to use and recharge them. Moreover, wearable devices might cause feelings of burden and discomfort. On the other hand, optics-based systems, such as computer vision and infrared, have line-of-sight detection problems (i.e., they cannot detect people behind obstacles), as well as privacy-related issues and overhead costs [13].

A wireless technology that uses electromagnetic waves (i.e., radar) to continually measure gait parameters at home or in a hospital, without a clinician's involvement, has been proposed as a suitable solution for many of the issues discussed above [10], [16]–[21]. The use of a radar system is appealing due to its reliable functionality in different lighting levels, protection of privacy, penetrations through obstacles, and long-range detection capabilities [22]. Radar sensors could make it possible to monitor and analyze gait outside the laboratory and capture information about the human gait and activity level during the person's everyday activities. It should be mentioned that there is little research about the radar's accuracy and applicability, and people may not feel comfortable installing radars in their place (since it is new and not a common technology), but there is growing interest in the use of radar systems in everyday life [23], [24].

Studies on the application of radar technologies in human gait assessment have been conducted to (1) analyze and obtain gait parameters [10], [16]–[21] and (2) recognize humans from their gait patterns [25]–[30]. For the first one, various radar signal processing methods have been proposed to extract gait characteristics such as speed, cadence, stride length, etc., while for the latter, machine learning and artificial intelligence (AI) have been deployed. However, the focus of this paper is to integrate machine learning algorithms with radar signal processing to identify the type of in-home activity performed by a subject and to detect in-home walking cycles to distinguish them from other in-home activities.

## II. STATE OF THE ART AND PROPOSED IMPROVEMENTS

Most of the current gait analysis systems are based on continuous-wave (CW) radars, while such systems enable Doppler/ micro-Doppler measurement but prevent range estimates [26], [31]–[34]. As shown in [35], one of the main drawbacks of CW radars is the effects of the approach angles of motion on the micro-Doppler patterns. For instance, a micro-Doppler signature of a person walking towards the radar at a relative 90º angle is different from that at 0º angle [19], [20], [25]. Previous works on developing radar-based gait detection systems have primarily focused on straight-line walking cycles to tackle the dependency of micro-Doppler patterns on the angle of motion [19], [20], [25]. However, the chance of in-home straight-line walking is very low as people walk randomly in their living environment. To overcome the dependency on the relative angle between the radar and a walking person, one possible solution is to obtain the walking speed through the changes in the position of the subject over time (i.e., velocity = position/time). A multiple inputs multiple outputs (MIMO) frequency modulated continuous wave (FMCW) radar [22], [30], [36] can provide the position of targets in addition to the micro-Doppler information, which makes it a good candidate for in-home gait monitoring assessment and activity recognition application [10].

Although, the speed of random walking could be extracted using a MIMO FMCW radar, the position of a subject performing other in-home non-walking activities such as vacuuming also changes over time. Therefore, for an in-home gait monitoring assessment and activity recognition system, we need an algorithm to distinguish walking periods from other activities and movements. In this paper, we propose a novel cloud-based in-home free-living activity recognition and gait monitoring system that integrates radar-based signal processing methods with a sequential deep learning algorithm to create an autonomous in-home gait monitoring and activity recognition (AI-GM&AR) system. Our proposed AI-GM&AR system uses sequential deep learning to identify the type of in-home activities, to recognize the in-home activity, to detect walking cycles and to distinguish them from other in-home activities.

Most of the research on human gait analysis and activity recognition has been done in a simple, large, and low clutter environment with a constrained range and limited activity [20], [22], [25], [36], [37]. However, when someone walks randomly in a cluttered environment such as a usual home, his walking patterns are completely different from that straight-line walking in a large area [20], [22], [25], [36], [37]. As will be shown in this paper, identifying walking periods and recognizing the type of activity a person performs is a complicated task in such an environment using radar signal processing methods.

The primary purpose of our research is to perform in-home free-living daily activity recognition and gait periods detection using radar technologies to have a record of subject's activity level and gait patterns during daily life activities. We perform gait and activity recognition studies in a familiar and commonly used environment, such as one's home. Moreover, unlike the work reported in [21] that used a complex radar system including four AWR1243 chips to create 192 channels to provide human point cloud information for 2D-DCNN, we used only one AWR1443Boost radar sensor. Note that for a real-time everyday application, we need a fast and simple algorithm, whereas an expensive high-resolution radar and complex signal processing are required to prepare point cloud information, as shown in [21]. Moreover, to make the system affordable, it is preferred to have fewer and inexpensive radar sensors. In this paper, we show that, without the need for an expensive high-resolution radar leading to complicated and computational-costly algorithms for detection, clustering, associations to extract point cloud information (x-y-z) [21], Joint Time-Frequency (JTF) representation of a human body obtained from a low-cost radar is reliable, rich and enough feature to be delivered to a sequential deep learning algorithm to be trained and so to predict in-home activities. In addition to the simplicity, compared to the point cloud information of the subject, another advantage of JTF patterns is that only one single transmitter and a receiver can provide spectrograms leading to a less expensive system.

Furthermore, we show that deep Gated Recurrent networks (GRU) [38] can extract temporal characteristics of the radar data and thus achieve sufficient recognition accuracy with relatively low complexity compared to the existing 2D-CNN methods [39]. Since it is very simple and fast preprocessing to get JTF patterns, all signal processing pipelines are performed in a low-cost standalone IoT Edge device without a need for allocating an extra laptop or PC to run the signal processing. In this paper, we utilize a Raspberry Pi to process radar raw data and perform all signal processing to be delivered to the GRU network. With a simple and fast preprocessing to create JTF patterns performed in a Raspberry Pi, streamed data is sent to the Cloud (Microsoft Azure), and the GRU network is applied to the streamed data to identify the type of activity a person is performing in real-time.

Our proposed cloud-based AI-GM&AR system not only detects walking cycles and captures gait values but also contains rich information about a person's daily activity level, such as the time when the person started and stopped walking, the distance of walking, how long the subject was stationary, how long the subject was active during a day (all other movements in addition to walking), if the target left home, etc. Additionally, the proposed standalone AI-GM&AR system integrated with a Raspberry Pi provides a record of the subject's activity level over time, washroom frequency, sleep/sedentary/active/out-of-home durations in the designed business intelligence dashboard developed in the Azure. These daily reports provide the level of activities of daily living used as an indicator of a person's functional status. The reports also could be used to collectively describe fundamental skills required to independently care for oneself [40].

The rest of this paper is organized as follows: (1) we initially describe the AI-GM&AR system design and the proposed algorithm for in-home activity recognition and gait monitoring in Section II; (2) in Section III, we present some samples of human radar signatures; (3) the proposed algorithm for the AI-GM&AR system is explained in Section IV, lastly (4) we discuss the experimental results in Section V.

## III. AI-GM&AR SYSTEM DESIGN

The diagram of our proposed AI-GM&AR system is presented in Fig.1, where the main components of the system include three parts: Client Side, Cloud and User Interface. To provide a detailed representation of the subject's daily activity, we focus our attention on the living room as this is the main area of the house where the subjects spend time and perform most of their activities, followed by the bedroom (to record the

---

**Algorithm 1:** In-Home Status Recognition Algorithm

**Input**: Radars Raw Data from each Single Board
**Output**: Activity Reports

**while** True:
  chirp=capture_raw_data ()
  room=PAD (chirp)
  **if** room == "in_bed"
    save_in_bed_date_and_time ()
  **else if** room="in_washroom"
    save_in_washroom_date_and_time ()
  **else if** room="out_of_home"
    save_out_of_home_date_time ()
  **else if** room=="Livingroom"
    status=check_status_of_livingroom (chirp)
      save_status_of_livingroom (status, date, time)

---

sleeping time and duration) and the washroom (to record washroom frequency, enter, exit, and duration). Therefore, to enable tracking a subject in the three main living areas, we installed a standalone system (a radar integrated with a single board) at the subject's bedroom, living room and washroom.

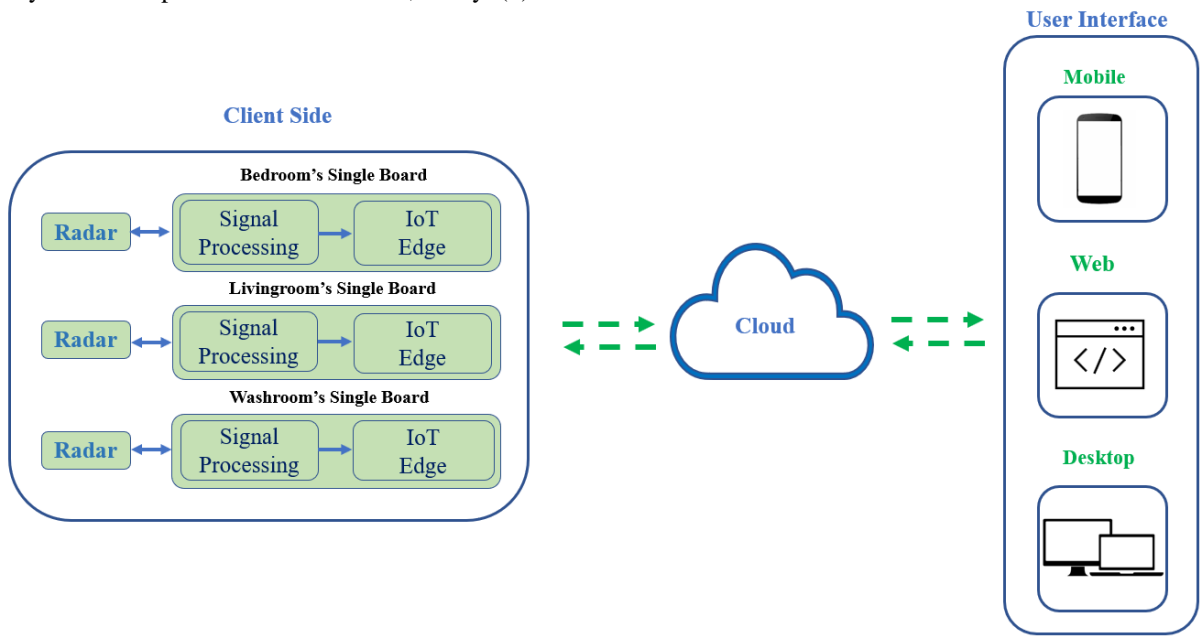

**Fig. 1.** Diagram of the proposed AI-GM&AR system. Three standalone units are installed at the subject's living environment collecting stream data and sending it to the cloud. In addition to gait parameters and the subject's current status, the subject's daily activity reports are recorded and shown using three different platforms: mobile, web and desktop apps.

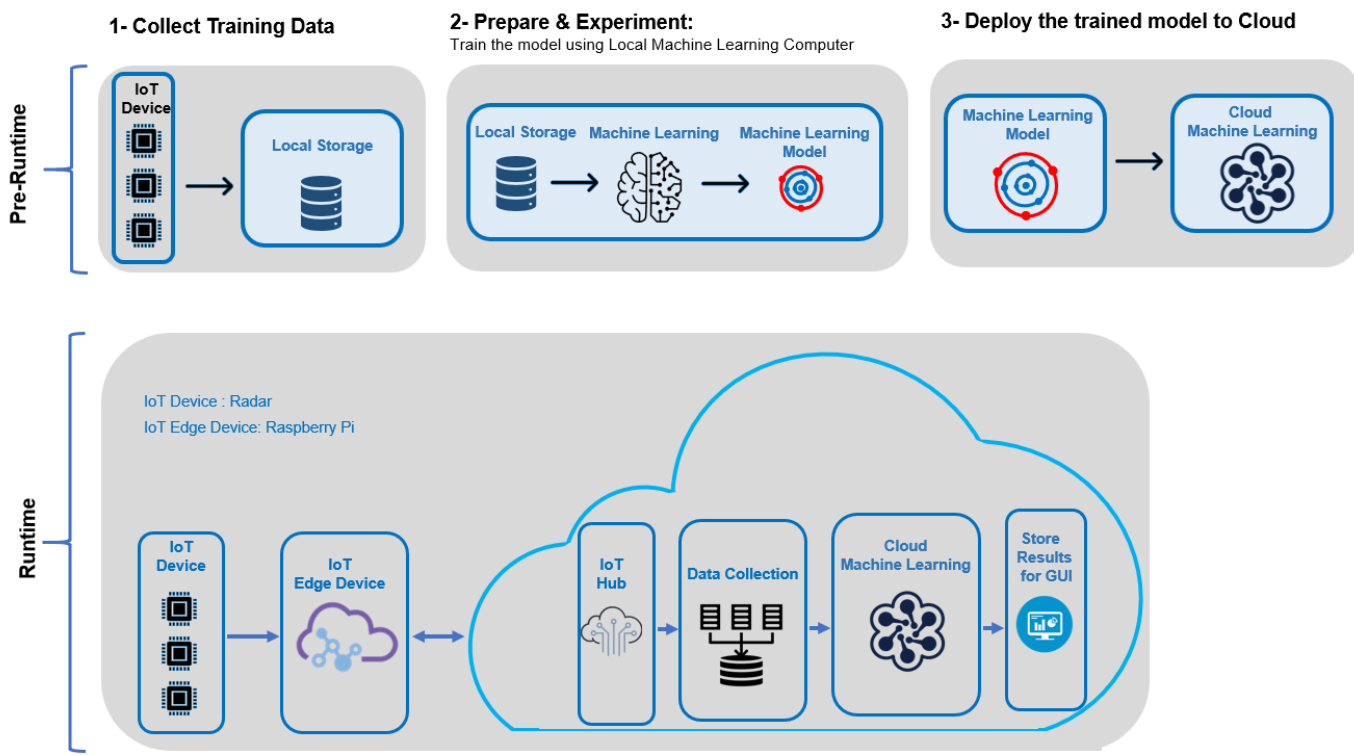

**Fig. 2.** Flowchart of the IoT-based AI-GM&AR system showing the pre-run time and run-time processes in the cloud.

Each system is used to send the radar configuration commands to run the radar, to store received raw signals, to preprocess the raw data, and then to transfer it to the cloud. As shown in Fig.1, each system in each room performs signal processing to detect the presence or absence of the subject. To identify which room is currently occupied, a Presence-Absence Detection (PAD) algorithm is applied to the radar raw data [41]. We refer interested readers for more detail of the PAD algorithm to our previous paper which was published in [41]. The PAD algorithm identifies the rooms as occupied or vacant. Through the IoT Edge, the data from the room occupied by the subject is sent to the cloud. Since our focus is on activity recognition and gait cycle identification in the living room using sequential deep learning, if the living room is occupied, then the joint time-frequency patterns of the subject will be sent to the cloud for further analyses. The pseudo-code of our proposed AI-GM&AR system is provided in Algorithm 1. As shown, the radar real-time raw data captured in each room is the system's input to generate the output of the subject's activity report. If the PAD algorithm identifies the presence of the subject in the bed (occupied bed), the time duration of in-bed status will be stored in the cloud database to record the sleep or in-bed time. On the other hand, if the presence of the subject is identified by the PAD algorithm in the washroom (occupied washroom), the entrance/exit time and the time duration the subject spends in the washroom will be recorded. Ultimately, if the PAD algorithm detects the subject in the living room, the JTF pattern will be sent to the cloud for further analysis. Deploying deep learning in the cloud, the type of activity performed by the subject will be predicted. If the PAD algorithm identifies the absence of the subject in all three areas, the status of out-of-home will be identified (a vacant room). The time duration that the subject spends out of home would be recorded.

*A. Deploying Machine Learning in Cloud*

There are two main steps in cloud computing to deploy real-time machine learning to recognize the type of activity and identify gait cycles in the living room using sequential deep learning [42]: pre-runtime and run-time processes. As shown in Fig. 2, in the pre-runtime step, we collect data from the IoT device (the radar sensor) to train a deep learning network. The model is trained and optimized in a local machine. The model is then deployed into the cloud to be used in the run-time section.

In the run-time step, radar sensors paired with Raspberry Pis (standalone sensors) are used to capture and preprocess streamed data from the environment and then send it to the cloud for further analyses. If the occupied room is the living room, the stream data is then transferred to the cloud and fed into the deep learning network to identify the type of activity and gait cycles [42]. More details are provided in Section IV.

*B. Representation of Human Radar Signatures*

Since the human body is non-rigid and human locomotion, including gait cycles, is a complex motion, the velocity of each segment of the human body performing different tasks varies over time [26], [31]–[34]. This results in various Doppler shifts (micro-Doppler) in scattered signals over time. Contrary to most of the current radar-based gait analysis systems using a CW radar that can monitor velocity but are unable to measure distance [26], [31]–[34], FMCW radars can overcome this shortcoming [22]. We use an FMCW radar to provide range-azimuth and micro-Doppler signatures of a person's in-home activities simultaneously. In this section, we cover the FMCW radar signal model to provide the JTF pattern of the human non-rigid body to be used as inputs for deep learning networks.

*1) FMCW Radar Signal Model*

In an FMCW radar, having "up chirps" (i.e., only positive slope chirps) for a transmitted signal s(t), the received signal at $l^{st}$ antenna element reflected by the target $x_l(t)$ can be modelled as (assuming that the target is a single point target):

$$x_l(t_f, t_s) = b_l \exp[-j(2\pi f_b t_f + 2\frac{vt_s}{\lambda_{max}} + \tau_l + \alpha_l \\ + \Delta \psi_l(t_f, t_s))] + e_l(t_f, t_s),$$  (1)

where $t_f$ and $t_s$ are fast and slow time indexes, $b_l$ and $\alpha_l$ are the channel's mismatched magnitude and phase, and $f_b$ is the beat frequency. The beat frequency is the frequency of an (IF) signal produced by an object in front of the radar located at the range of R calculated by

$$f_b = S\frac{2R}{c}$$  (2)

where S is the rate of increase in the frequency of the sinusoid, and c is the speed of light in free space. Moreover, $v$, $\lambda_{max}$, $\tau_l$, $\Delta\psi_l(t_f,t_s)$, and $e_l(t_f,t_s)$ in (1) are the target's radial velocity, the wavelength corresponding to the start frequency of the FMCW ramp, the phase shift at $l^{st}$ receiver due to the angle of arrival (AoA), the residual phase noise and the additive noise, respectively.

The range of an object is limited by IF bandwidth supported by the radar device written as:

$$R_{max} = c\frac{f_{bmax}}{2S}$$  (3)

where $f_{bmax}$ is the maximum supported IF bandwidth. Since the $f_{bmax}$ is also dependent on the ADC sampling frequency, an ADC sampling rate of $F_s$ limits the maximum range of the radar to

$$R_{max} = c\frac{F_s}{4S}.$$  (4)

Since the range estimation of a detected target in the FMCW radar is primarily dependent on the frequency resolution of the Discrete Fourier Transform (DFT) performed on the base-band signals, the new range resolution of the radar is given as

$$\Delta R = \frac{c}{2B}.$$  (5)

In an FMCW radar, to resolve objects in range, the Fast Fourier Transform (FFT) is performed on the beat signal (range-FFT) that provides the relative radial distances (i.e., range) of various objects scanned by the radar. This is done such that the frequency of the peaks in the range FFT directly corresponds to the range of the target. Moreover, to obtain the velocity information of an object, a sequence of chirps separated by Tc (Tc =τ+T) called a frame is required.

The range-FFTs corresponding to each chirp will have peaks in the same location but different phases. The measured phase difference (ω) corresponds to a motion of the object of $v \cdot T_c$. Therefore, performing the second series of FFTs (Doppler-FFT) across the chirps, the velocity of objects can be calculated by

$$\omega = \frac{4\pi v T_c}{\lambda} \Rightarrow v = \frac{\lambda \omega}{4\pi T_c}.$$  (6)

To calculate the unambiguous range of velocity, the phase of the signal should be less than π, which is written as:

$$|\omega| < \pi \Rightarrow v < \frac{\lambda}{4T_c}.$$  (7)

Thus, to detect higher velocities, we require closely spaced chirps (shorter Tc). The velocity resolution ($v_{res}$) is inversely proportional to the frame time ($T_f$) or the number of chirps (N) per frame given by

$$v_{res} = \frac{\lambda}{4T_f}, T_f = NT_c.$$  (8)

*2) Joint Time-Frequency Pattern*

As shown, FFT could be used to obtain the frequency spectrum (i.e., the velocity of a target), reflecting the signal features in the frequency domain. However, for human activity detection, each segment of the body creates frequency shifts; for example, during walking cycles, the torso and limbs have different motion velocities and acceleration, thus generating time-varied Doppler frequency shifts. Consequently, radar-received signals from human activities are non-time-stationary in the time domain. Therefore, the important and informative time-varying micro-Doppler frequency information will be lost if we perform FFT on the scattered signals from a human body. However, Short-Time Fourier Transform (STFT) is a technique to overcome the issue and to represent the time dependency in the Fourier transform (spectrogram) [33], [43].

Here, we use STFT to get the time-frequency distribution of human activity signals. The STFT can be defined as:

$$\text{STFT}(t, f) = \left|\sum_{n=0}^{P} w(n) x(t-n) e^{-j2\pi fn}\right|^2$$  (9)

where, w(n) is the short-time analysis window function, and x(t) is the signal to be transformed.

In the FMCW radar, however, since reflections from a human body occupy multiple cells of range bins calculated in the range profile, STFT should be applied to all occupied bins. In this paper, since our target is a single subject, we perform STFT to all range bins. This simplicity helps us avoid other signal processing such as detection (to capture occupied bins), clustering to cluster the detected bins, and then association to associate the new bin to the previously occupied bin; these are required processes for multiple people detection. To do so, coherent accumulation is firstly performed on range samples and next on the channels vector to increase the signal intensity [41]. In fact, we benefit from the MIMO feature of our MIMO FMCW radar to increase the signal intensity and improve detection [41]. Moreover, in another step, we use this feature to provide azimuth information of the subject [44].

Consequently, the JTF pattern of reflected signals of a human performing various activities is calculated as

$$\text{JTF} = \sum_{i=1}^{L} \sum_{j=1}^{K} \text{STFT}$$  (10)

where L is the number of channels and K is the number of range bins. The resulting output of the JTF is a three-dimensional plot representing the frequency content over time of the signals reflected from a human body [41].

## IV. PROPOSED AI-GM&AR ALGORITHM

The block diagram of our AI-GM&AR algorithm flowchart is illustrated in Fig. 3. As shown, the proposed algorithm consists of two separate processes: walking periods identification/activity recognition and gait extraction. In this paper, we cover the method of in-home walking period identification and activity recognition.

In our proposed system, radar raw data is collected from a MIMO FMCW radar. As shown in Fig. 3 and Algorithm 2, range FFT is applied to the received chirp samples from the FMCW radar to obtain the range information, and then the range-profile of all channels is created. To remove leakage from the transmitter antennas to receiver antennas, mutual coupling reduction is applied. Then, a stationary clutter removal algorithm is applied to the range-profile to remove signals reflected from stationary clutters. The average value of the signal is computed and subtracted from the aggregated signals; removing the average is equivalent to eliminating the stationary scatters [41]. Therefore, the only remaining signals in the range profile are from humans, which are caused by chest motions (breathing) and other motions created by performing different activities.

**Algorithm 2: In-Home Activity Recognition Algorithm**

check_status_of_livingroom (chirp):
**Input**: Radar Raw Data in Livingroom (Radar Unit #1)
**Output**: Type of Activity in the Living Room

```
chirp=capture_radar_raw_data_from_livingroom ()
fft=fast_fourier_transform (chirp)
complex=fft[0:int (fft.size/2)]
mcr=mutual_coupling_reduction(complex)
cr=clutter_removal (mcr)
data=save_in_database (cr)
jtf=joint_time_frequency (cr)
result=GRU (jtf)
if result=="empty"
 return "empty"
if result=="sedentary"
 return "sedentary"
if result=="washing"
 return "washing"
if result=="vacuuming"
 return "vacuuming"
if result=="in_place_movement"
 return "in_place_movement"
if result=="walking"
 return gate_extraction (data)
```

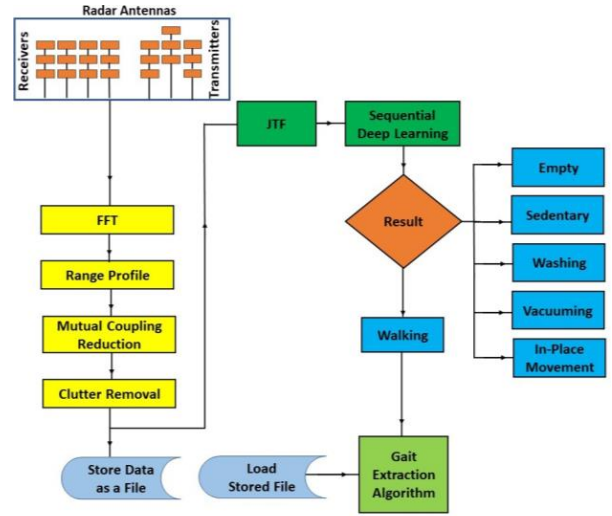

**Fig. 3**. AI-GM&AR System Flowchart. Firstly, features for sequential deep learning will be provided and delivered to GRU. If walking cycles are identified, a gait extraction algorithm will be applied.

In order to capture human in-home gait patterns, we propose our AI-powered system. Since most of the current gait parameter extraction methods are based on micro-Doppler patterns [10], [45], there is a pressing need to identify and discriminate walking cycles from other movements. Without identifying walking periods, other activities creating micro-Doppler shifts in the JTF pattern (10) might be identified as walking cycles while they are not the actual walking cycles [22]. The same issues arise for those methods proposed to extract walking speed based on the change of the subject's position. For example, it is essential to identify walking cycles in order not to calculate other in-place movements as walking cycles. Hence, any in-home gait extraction method is prone to failure if the system is not intelligent to identify human's in-home activities and discriminate between them. Therefore, to have precise and accurate gait data as well as to have a track of a person's in-home activities over long periods, the system should be able to identify the type of in-home activities a subject performs. As shown in Fig. 3, we deploy sequential deep learning networks to identify walking cycles and to recognize other in-home activities. Since the JTF pattern of a human body is composed of rich information of human activities, JTF patterns of the subject are calculated and delivered to sequential deep learning algorithms. Six classes are defined: (1) "Empty", (2) "Sedentary", (3) "Washing", (4) "Vacuuming", (5) "In-place movement", and (6) "Walking". These are some essential activities each person performs during the day. Algorithm 2 shows the pseudo-code of the proposed activity recognition algorithm. As shown, the deep learning network would predict the type of activity a person is performing and send out the result as the current status. For the case of detected walking periods, a gait extraction algorithm will be applied to the stored range-profile samples to obtain gait values. In our future publications, we will explore the methods used for in-home gait parameters extraction.

*A. Deep Learning for the AI-GM&AR System*

As mentioned, our system uses JTF patterns of activities to recognize in-home human activities, as well as to identify walking intervals. However, complex signal processing is required to map the JTF patterns to human's specific activity, which is mathematically not tractable [46]. For this reason, we are motivated to adopt machine learning as an effective tool for our AI-GM&AR system. In our previous work [47], we used a Random Forest classifier to identify walking cycles from other in-place movements. However, conventional machine learning algorithms are limited in their capacity to fully capture the rich information contained in complex data, particularly time-varying samples [46]. Our proposed system in this paper leverages deep learning approaches [48] to use the resulting time-varying signatures of the subject being monitored. The use of multiple deep layers in a single network enables not only the efficient extraction of a subject's features but also the building of a classification boundary [46].

Many deep learning models have shown exceptional promise in radar-based human activity recognition systems [25]–[30]. Commonly, the raw data is generally converted into a 2-D spectrogram using the STFT method while being treated as an optical image. In these systems, the corresponding architectures such as 2-D convolutional neural networks (2D-CNNs) are used. However, since a human body motion consists of a series of associated postures through time, ignorance of temporal characteristics leads to a complex network with a huge number of parameters but limited recognition accuracy. For this reason, CNNs are not a good algorithm for such time-varying data.

On the other hand, deep recurrent neural networks (DRNN) have successfully addressed classification problems that feature temporal sequences [48]. DRNNs use a hidden node that serves as memory, passing previous information to the next state for processing sequential inputs. Through this process, a DRNN can extract the temporal features of data. Long short-term memory (LSTM) and gated recurrent unit (GRU) are the two common models for sequential learning [46]. However, because of the complex structure of a single LSTM unit, the LSTM network contains many parameters and so requires a larger sample size. LSTM contains three gates: the forget gate, the input gate, and the output gate.

On the contrary, a GRU network has a simpler structure and fewer parameters. A GRU network includes only two gates: the reset gate and the update gate. From a spatial complexity perspective, LSTM has more parameters than GRU, therefore GRU has fewer computation costs than LSTM [38]. In this paper, we show that JTF patterns have enough features for a single subject in-home monitoring, and the GRU network is a promising algorithm to be used for time-varying signatures of human activity classification. We demonstrate that GRU can extract temporal characteristics of the radar data and thus achieves sufficient recognition accuracy with relatively low complexity without the need for the subject's point cloud information [39]. Advantages of JTF patterns, compared to the point cloud information, are that such a system can provide spectrograms using only one single transmitter and a receiver. Additionally, preprocessing is faster and simpler. An alternative approach using point cloud information would require an expensive high-resolution radar and complex signal processing.

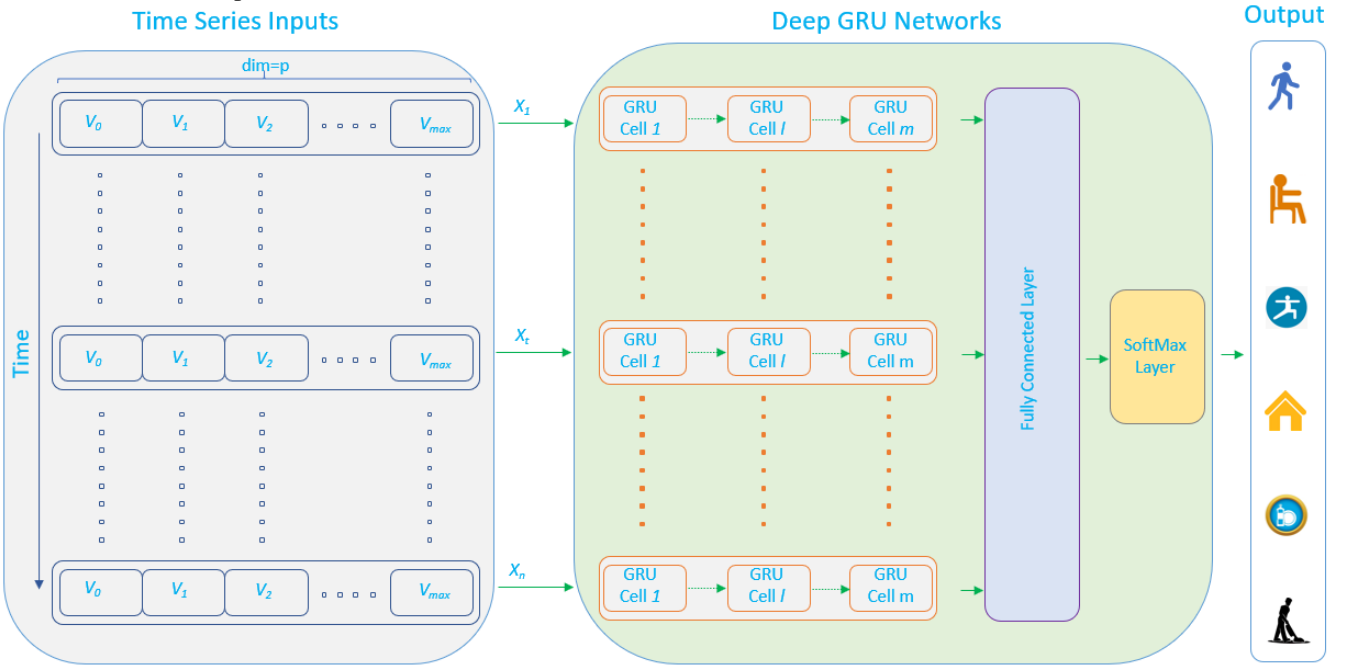

**Fig. 4.** Schematic of deep GRU networks; architecture of a deep GRU network with m GRU layers and fully-connected layers for sequence-to-sequence modeling. The network outputs are the subject's living room status: "Empty", "Sedentary", "Washing", "Vacuuming", "In-place movement", and "Walking".

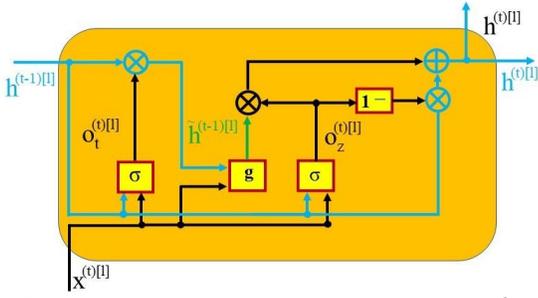

**Fig. 5.** Architecture of a typical GRU cell of the $l^{th}$ layer at time t.

*1) Proposed GRU Networks Architecture*

Fig. 4 illustrates the architecture of the deployed deep GRU networks. The network consists of multiple hidden layers, including the inputs (the time-series data of length n), GRU cell layers, fully connected layers that combine all the features learned by GRU layers for classification, and the output layer. Therefore, the output size is equal to the number of classes. The SoftMax layer normalizes the output of the former layer to be used as the classification probabilities. As shown in Fig. 5, each GRU layer has GRU cells, including an independent set of weights and biases shared across the entire temporal space within the layer.

The GRU cell consists of inputs, a hidden state, and two gates: a reset gate and an update gate. The hidden state memorizes the cell state at the previous time step while the reset/update gates are derived using both the previous time step and the input data at the current time step. Controlled by the reset/update gates in each GRU cell, valuable information along the temporal sequence can selectively be extracted and propagated to capture the long short-term time dependence in a time-varying system.

At the time step $t$ ($t=1, …, n$, where n is the total number of time steps), with the input of $X^{(t)[l]}$ of the $l^{th}$ hidden layer, the reset gate vector is calculated as [38]:

$$o_r^{(t)[l]} = \sigma\left(W_{rh} h^{(t-1)[l]} + W_{rx} X^{(t)[l]} + b_r^{[l]}\right). \quad (11)$$

As shown in Fig. 5, the input data $X^{(t)[l]}$ represents the micro-Doppler pattern of the subject at time $t$ with the dimension of p covering the subject's created Doppler velocity range of $v_0 = -\frac{\lambda}{4T_c}$ to $v_{max} = \frac{\lambda}{4T_c}$.

In the second step, a non-linear activation *tanh* function will be applied to the result of an element-wise multiplication (Hadamard product) of the previous hidden state with the reset gate vector multiplied by a trainable weight and at the same time summed with the current input multiplied by a trainable weight.

$$\tilde{h}^{(t)[l]} = g\left[W_{\tilde{h}x}\left(o_r^{(t)[l]} \odot h^{(t-1)[l]}\right) + W_{\tilde{h}x} X^{(t)[l]} + b_{\tilde{h}}^{[l]}\right]. \quad (12)$$

The next step is to create an update gate vector, the process of which is similar to create the reset gate vector but with its unique trainable weights.

$$o_Z^{(t)[l]} = \sigma\left(W_{Zh} h^{(t-1)[l]} + W_{Zx} X^{(t)[l]} + b_Z^{[l]}\right). \quad (13)$$

An element-wise multiplication of the reset gate vector with the element-wise inverse version of the update gate vector is taken. Lastly, the result is summed with the Hadamard product of the update gate vector with the previous hidden state. Consequently, a new and updated hidden state will be calculated to be used in the next GRU cell.

$$h^{(t)[l]} = o_Z^{(t)[l]} \odot h^{(t-1)[l]} + \left(1 - o_Z^{(t)(l)}\right) \odot \tilde{h}^{(t)[l]}. \quad (14)$$

Note that in the above equations, $l$ represents the $l^{th}$ hidden layer, $X^{(t)[l]}$ is the input of the $l^{th}$ hidden layer, σ and *g* are activation functions, W and b denotes weights and bias, respectively, $h^{(t)[l]}$ represents the hidden layer of the next GRU cell, the letter r and Z indicate the reset gate and the update gate, respectively. ⊙ denotes an element-wise multiplication (Hadamard product).

*2) Network Inputs*

As mentioned, time-varying information of the subject's spectrogram is considered inputs for the GRU networks. Fig. 6 shows the network inputs in detail. After optimizing the network parameters, at the time step *t*, $X^{(t)[l]}$ is a vector of $1 \times 256$ consisting $τ= 25$ *m*s of a subject's micro-Doppler signature derived from (10). The total number of time steps is set to 50. Note that for the time steps, we observed that when it is greater or less than 50, the performance of the GRU networks degrades. Therefore, the dimension of the input of the GRU networks is (1, 50, 256); in total, t=25×50=1250 *m*s of the subject's spectrogram is fed to the GRU networks.

*3) GRU Network Implementation*

Since the performance of deep learning networks highly depends on their hyper-parameters to control how it learns the training dataset, it is necessary to optimize the hyper-parameters utilized by each network. To find the best hyper-parameters for the GRU network, a range of values was specified for five crucial hyper-parameters: learning rate, batch size, activation function, optimization algorithm and number of layers. A grid search was conducted to obtain the hyper-parameters, where the optimized hyper-parameters for the GRU networks are summarized in Table I.

We implemented the proposed GRU networks in TensorFlow (Keras) with a cross-entropy loss function in 200 epochs.

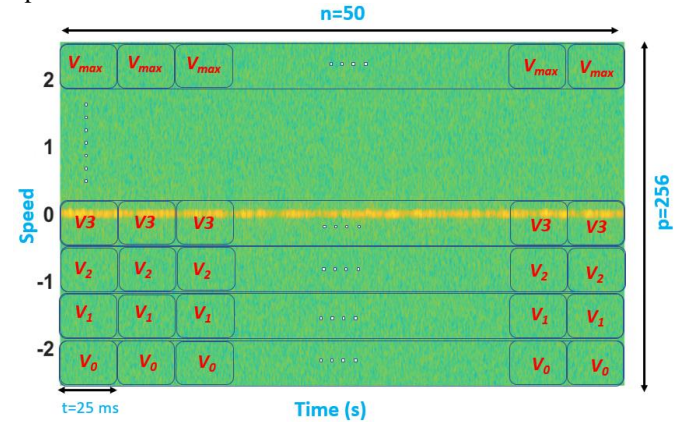

**Fig. 6.** Time series inputs for GRU networks; inputs are a vector of (1, 50, 256).

Table I
GRU NETWORKS HYPER-PARAMETERS

| Hyper-parameters | Optimized values |
|---|---|
| Activation function | Rectified Linear Unit |
| Batch size | 512 |
| Number of layers | 7 |
| Learning rate | 0.01 |
| Optimization algorithm | Adam |

## V. EXPERIMENTAL RESULTS

### A. Our Devices

Since it is very simple and fast preprocessing to get JTF patterns, all signal processing pipelines are performed in a low-cost IoT Edge device without a need for allocating an extra laptop or PC to run the signal processing. As shown in Fig.1, on the client-side, we use three radars integrated with Raspberry Pis 4 (Model B, RAM 2GB) in the subject's bedroom, living room and washroom separately. We use a Raspberry Pi to process radar raw data and perform all signal processing to be delivered to the GRU network. With a simple and fast preprocessing to create JTF patterns performed in a Raspberry Pi, streamed data is sent to the Cloud (Microsoft Azure), and the GRU network is applied to the streamed data to identify the type of activity a person is performing in real-time.

For each room, only one AI-powered IoT- based standalone system is used. Our radar sensors are mm-wave FMCW radar systems (from TI Co. Ltd) [49]. Our system uses only one AWR1443Boost radar for each room. Since we use the time-varying signature of a human body, one single radar provides sufficient and rich information of each type of activity without the need for a complex and expensive system to provide point cloud information.

As shown in Fig. 7, the radar system has three transmitters ($T_1$, $T_2$, and $T_3$) and four receivers ($R_1$, $R_2$, $R_3$ and $R_4$), allowing the construction of a 12 virtual receiver array for each chip. Note that we use the virtual receiver array to provide azimuth information of a subject to find the subject's position for future analysis. We refer interested readers for more detail of extracting azimuth information of a subject to our previous publication where we used a Capon algorithm [44].

### B. Radar Configuration

In this experiment, we use the following parameters for radar configuring: chirp duration: 90 μsec, chirp slope: 43.03 MHz/μsec, chirps per frame: 256, frame period: 98 msec, frequency bandwidth: 3870 MHz, operating frequency: 77 GHz, and A/D sampling rate: 3400 ksps. Using this chirp configuration, the following parameters are obtained: maximum detachable range: 5.9238 m, range resolution: 3.8733 cm, velocity resolution: 0.02 m/s, maximum velocity: 2.5414 m/s. Hence, p in Fig. 5 and Fig 6 is set to 256.

Since we consider only one subject to be monitored, we use the JTF pattern as (10), which has the time-varying information of the whole segments of the human non-rigid body. To generate the JTF signature (10) of each activity, we choose a small Hamming window for time-frequency analysis of the real-time radar human activities /gait signals. The length of the Hamming window is 128 samples. We perform a Fourier Transform on the signals in the time window for each time instant "t" to obtain the time-frequency spectrogram.

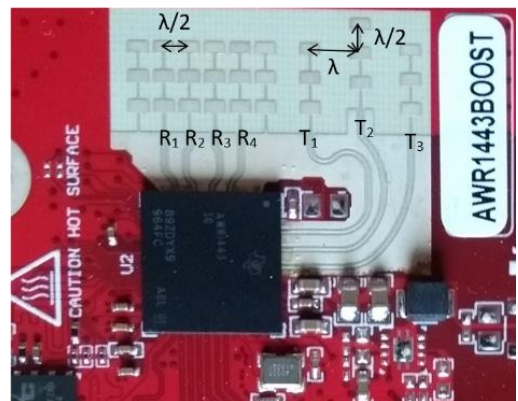

**Fig. 7.** The AWR1443Boost PCB antenna.

### C. Dataset Construction

Most of the current research for human gait analysis and activity recognition was done in a simple, large, and low clutter environment with a constrained range and limited activity. However, if we walk in a cluttered environment such as a usual home randomly, walking patterns are completely different from that straight-line walking in a large area. As we demonstrate in this article, identifying walking periods and recognizing the type of activity a person performs is a complicated task in such an environment. Since our goal is to provide an in-home gait cycles identification and activity recognition system, we collected data at two completely different environments: (1) a large environment without any clutter (Fig. 8) and (2) a highly-cluttered typical apartment. The purposes of these two different experiments are (1) to compare and show the challenges and effects of the cluttered environment on radar signals and (2) to show the reason for the need for an intelligent algorithm for in-home walking cycles (particularly non-straight-line walking).

Note that all samples were collected from two subjects. Due to COVID-19 restrictions, it was not possible to collect data from more participants.

#### 1) Experiments in a Large and Low Clutter Environment

To collect samples in a large environment, we defined four classes of activities: (1) "Empty", (2) "Sedentary", (3) "In-place movement" and (4) "Walking". Our policy for data collection is as followed:

(1) For the "Empty" class, no one was in front of the radar.

(2) For the "Sedentary" class, subjects were sitting in front of the radar with free and normal hand/head motions or working with a cellphone.

(3) For the "In-place movement" class, subjects were asked to mop the floor and pick up objects from a table.

(4) For the "walking" class, the subjects were invited to walk back and forth on the traces at five different angles in front of the radar, as shown in Fig. 8.

Time-frequency spectrograms (the JTF pattern calculated from (10)) of the signal of a subject walking at 0° and 60° are shown in Fig. 9 (a) and (b), respectively. The horizontal axis represents the time in the figure, and the vertical axis represents the Doppler frequency caused by body motion. Since the different parts of the human body do not move with constant radial velocity, the small micro-Doppler signatures of the human body are time varying. Therefore, STFT analysis techniques are an excellent tool to reveal more gait characteristics from time-varying gait signals. JTF patterns of the subject moping the floor and sitting in front of the radar are provided in Fig. 9 (c) and (d), respectively.

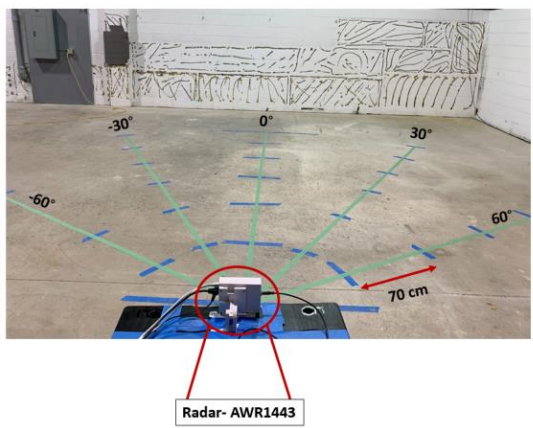

**Fig. 8.** Experimental setup for data collection in a low-cluttered environment.

As evident from the spectrograms of different activities, all type of human activities creates distinguishable micro-Doppler frequency at various frequencies. Although the cleaning and sitting patterns are completely different from the walking patterns, we need an algorithm to distinguish walking cycles from other activities. This is because most of the gait parameters extraction methods are based on the Doppler patterns of the gait cycles.

As mentioned, since Doppler-based algorithms depend on the angle of motion, one possible solution is to obtain the walking speed through the changes in the position of the subject over time. Even for this type of gait capturing method, we show that an algorithm is needed to distinguish walking cycles from other movements.

The trajectory of the subject walking at 0° and 60° relative to the radar is provided in Fig. 10 (a) and (b). The horizontal and vertical axes represent the x and y positions of the subject, respectively. The trajectory of the subject moping the floor and sitting in front of the radar are also shown in Fig 10 (c) and (d), respectively. As shown, the subject walked in a straight line over time while the position of the subject sitting and moping on the floor is clearly detectable. Although for the case of sitting, the subject was fixed at a specific position, the subject had displacement when moping the floor. Therefore, an algorithm is still required to distinguish between these displacements for the case of walking and other movements. As will be shown in the next section, this need is more evident for in-home monitoring. Therefore, for any gait monitoring algorithm and activity recognition, there is a strong need for an algorithm to discriminate between walking periods and other activities. To fulfill this requirement, we deploy the GRU algorithm, as discussed in Section V. A.

For the experimental setup shown in Fig. 8, with the defined scenarios, we collected 105 minutes samples in total, roughly 95 minutes for training and 10 minutes for test sets.

*2) Experiments in a real-world apartment*

As mentioned, datasets collected by other researchers in this field were collected in controlled situations in a large environment with no clutters [20], [22], [25], [36], [37]. While this type of initial research is critical in the path to developing practical systems, there needs to be research exploring how to computationally deal with subjects behaving naturalistically and with everyday objects seen in one's homes, as they would in a real home uncontrolled environment. Previous research reported in [21] explored a highly cluttered environment but in controlled scenarios using an expensive and complex high-resolution radar for creating a human point cloud information. However, there has been no reported research on a single sensor radar dealing with in-home gait and activity monitoring using the JTF patterns. To compile this dataset, we invited subjects to walk at their home completely randomly, without any predefined path to follow, at their selected speed, as well as perform various in-home activities that are very similar to walking but are not actual walking cycles. If we walk or move in a highly cluttered environment, we create various multipath effects that should be removed from the actual signals coming from the subject. This effect adds more complexity to the signal processing chain in addition to the required high resolution and complex radar sensors. To compile datasets in a usual home, we installed the radar sensor in a living room of a small apartment. We defined six classes: (1) "Empty", (2) "Sedentary", (3) "Washing", (4) "Vacuuming", (5) "In-place movement", and (6) "Walking". Our purpose for the "Empty" class was to identify the presence or absence of the subject in a living room. For the "Sedentary" class, the purpose was to know how long the subject was sedentary (not active) during a day. This information could help the subject have a track of his daily activity and modify his future activities. Moreover, since most of the available vital signs detection algorithms are effective for a stationary subject, knowing whether a subject is stationary gives insight into the efficacy and confidence of breathing rate and heart rate algorithm outputs [50]. For instance, if it is determined that a subject is moving, it may be best to withhold these estimates until the subject is stationary so as not to give erroneous results [50].

The purpose of defining the "Washing", "Vacuuming", and "In-place movement" classes was not only to distinguish such activities from gait cycles but also to have a record of a subject's daily activities that the subject is not stationary nor walking. As shown in Fig. 11, the JTF patterns of these activities are very close to in-home walking cycles, so they should be discriminated from each other. The "walking" class was defined since it is crucial to know the actual in-home walking cycles to perform gait extraction algorithms. Without identifying the actual gait cycle, applying gait extraction algorithm on other non-walking activities cause serious error in the gait data over time. Any tiny error in gait parameters would cause misleading information either in prediction or prevention [12], [51], [52]. Therefore, with the defined classes, our system provides unique daily reports of in-home human activities. The system collects gait data over time, stores the total time that the subject was active and the total time when the subject was stationary, and the type of activities the subject performs over a day.

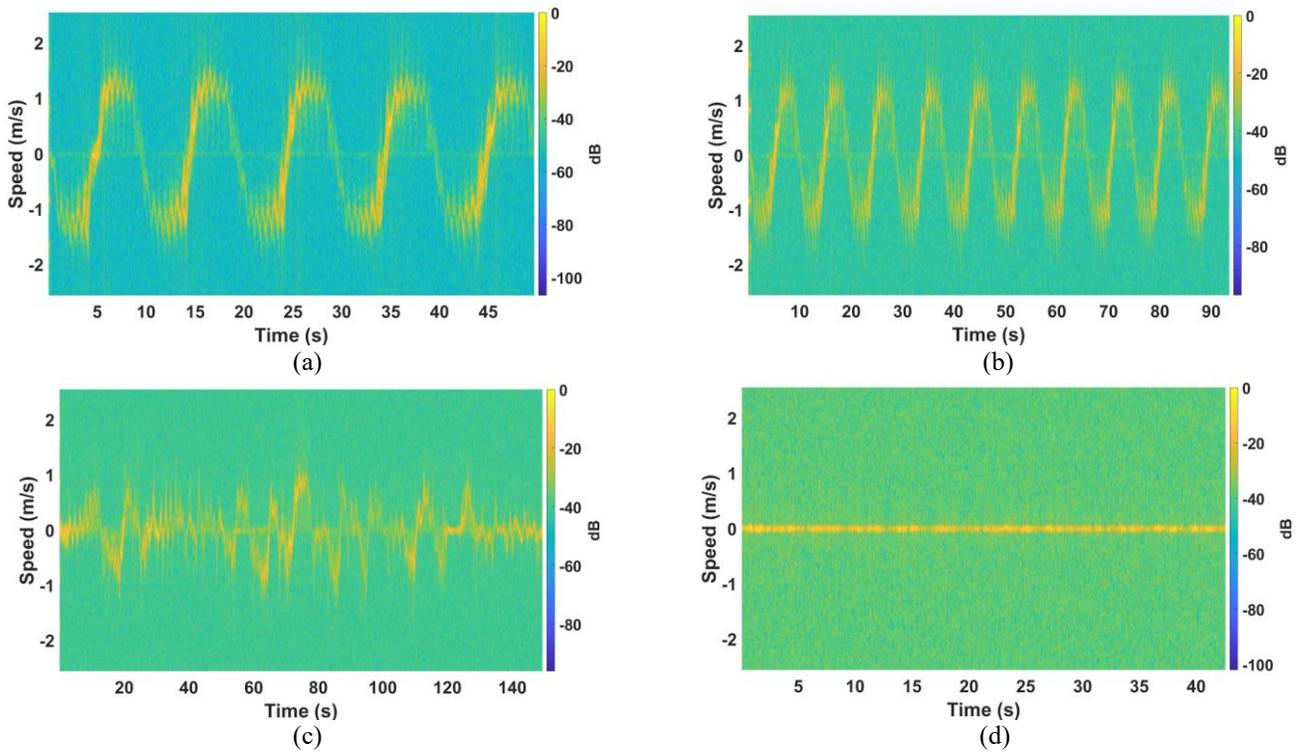

**Fig. 9.** Time series inputs: JTF patterns of the subject (a) walking in a straight line in front of the radar at 0 ° (b) walking in a straight line in front of the radar at 60 ° (c) mopping the floor and (d) sitting in front of the radar.

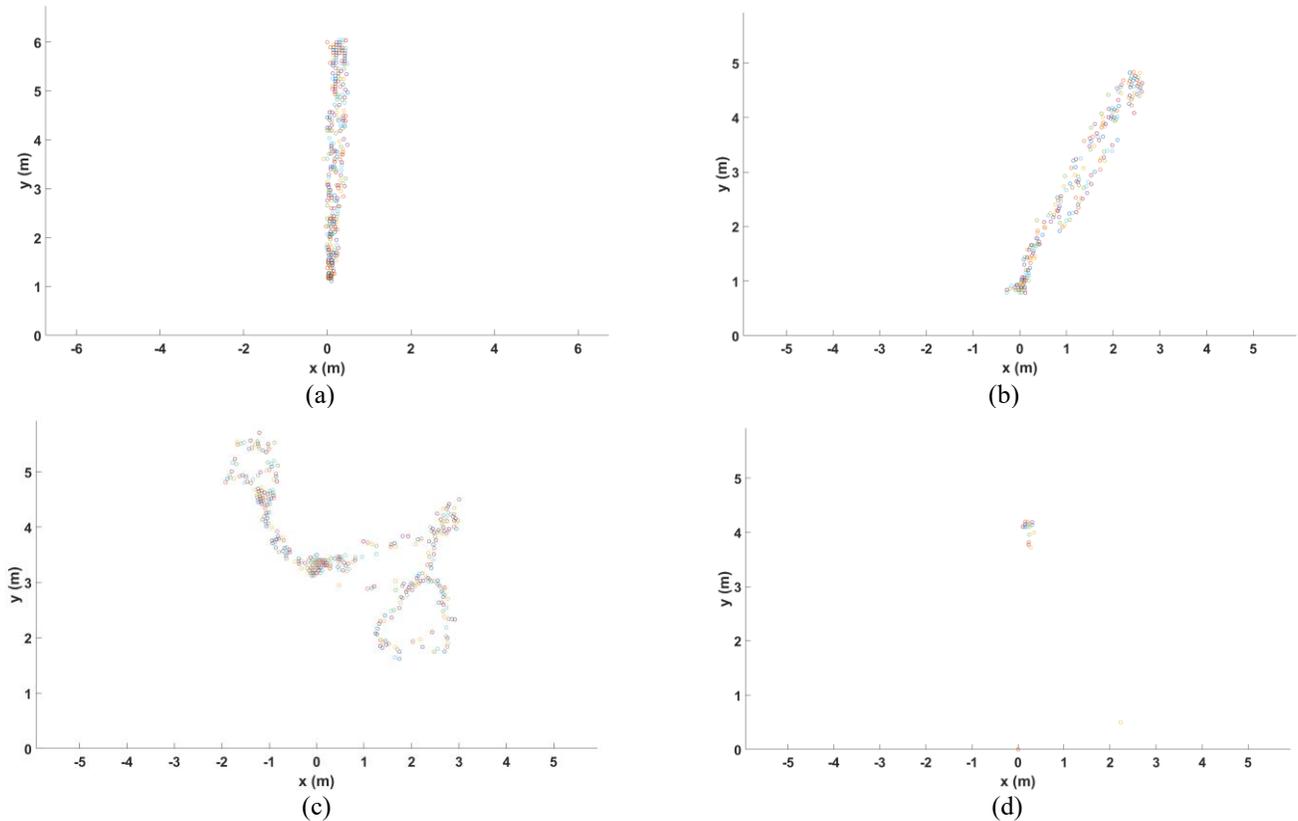

**Fig 10.** Trajectory of the subject (a) walking in a straight line in front of the radar at 0 ° (b) walking in a straight line in front of the radar at 60 ° (c) mopping the floor and (d) sitting in front of the radar.

This dataset was collected over five days to make a session-independent dataset (i.e., training data were collected over four days with the same subjects and tests data were collected on day 5 with the same subjects). This helps us to evaluate the network performance to predict new in-home activities for the same subjects over time. For dataset construction, two subjects performed their normal and natural daily life in their apartment.

These are the protocol we followed for in-home dataset construction:
1. For the "Empty" class:
   - There was no subject in front of the radar, and the room was not occupied by any alive body while windows kept open.
2. For the "Sedentary" class:
   - A subject was sitting still at different places on sofas.
   - A subject was sitting at different places on sofas while working with a cellphone.
   - A subject was sitting at different places on sofas while working on the laptop.
   - A subject was sitting at different places on sofas in the normal way (i.e., talking, moving his body, leg displacement, etc.).
   - A subject was sitting at different places on sofas facing back, toward the radar.
3. For the "Washing" class:
   - A subject was washing dishes in the kitchen.
4. For the "Vacuuming" class:
   - A subject was vacuuming the entire living room.
5. For the "In-place movement" class:
   - A subject was working out.
   - A subject was squatting.
   - A subject was picking objects from tables.
   - A subject was sitting and standing from a chair.
6. For the "Walking" class:
   - A subject was randomly walking in different directions at all parts of the living room, even behind desks and coffee tables at his own selected speed.

In total, we collected 200 minutes of data from each subject for training and 30 minutes for test sets. An example of the JTF pattern of in-home random walking is provided in Fig. 11 (a). As shown, unlike the straightforward straight-line walking JTF signature (Fig. 9 (a) and (b)), the time-varying signature of in-home walking is a complex pattern. Moreover, contrary to Fig. 9 that all patterns are distinguishable, we cannot clearly discriminate the patterns of different in-home activities. For instance, the JTF pattern of vacuuming (Fig. 11 (b)) is close to that of walking (Fig. 11 (a)). Other spectrograms such as washing and sitting patterns are shown in Fig 11 (c) and (d), respectively. As seen, for in-home activity recognition, we cannot easily map the JTF patterns to the subject's specific activity.

Furthermore, the trajectory of in-home walking is shown in Fig. 12 (a). As seen, the walking trajectory is completely random while it is organized in Fig. 9 (a) that the subject followed a traced straight line. Compared to the trajectory of the subject when vacuuming the living room, the subject changed his position significantly. Thus for gait extraction based on change of position, there is a strong need for gait cycle identification.

For in-home activity recognition and gait monitoring, we deliver JTF patterns of the subject doing different activities defined above (some examples are provided in Fig. 11) to train the deep learning network. The type of in-home activities will be trained and then predicted in real-time new scenarios.

*D. Results*

To train and test machine learning algorithms, a common approach is a K-fold validation method. For example, in a 5-fold validation method, 80% of data is used for training, 10% for validation, and the remaining 10% for testing [21]. This approach is acceptable for training/validation. However, to assess a deep learning network used for radar datasets, it is essential to have completely unseen samples for the test set to validate the network architecture in a real-life application. The reason is that although with the K-fold validation method, the samples sound unseen for testing, since the radar frame rate is very high, the results of several consecutive frames are very close to each other. Radar signals are not like camera images but, due to the higher frame rate, most of the data collected in a short amount of time are very similar.

Therefore, to avoid such issues, for the test set, we should collect samples that are either session-independent or collected from a complete new person. Note that, as defined, we regard the data collected on different days as session-independent datasets. To show the robustness and effectiveness of our proposed method for human activity recognition and gait cycle identification, we followed four procedures: 1. Train/validation set based on samples collected from two subjects performing various activities (defined in Section V. A .1) in a large low clutter area, and the test set based on session independent datasets from the two subjects (Scenario #1). 2. Train/validation set based on samples collected from subject "A" performing various activities (defined in Section V. A .1) in a large low clutter area, and the test set based on samples collected from subject "B" performing various activities in the same area (Scenario #2). 3. Train/validation set based on samples collected from two subjects performing various activities (defined in Section V. A .2) at their apartment (living room), and the test set based on session independent datasets from the two subjects (Scenario #3). 4. Train/validation set based on samples collected from subject "A" performing various activities (defined in Section V. A .2) at the apartment (living room), and the test set on samples collected from subject "B" performing various activities in the same room (Scenario #4).

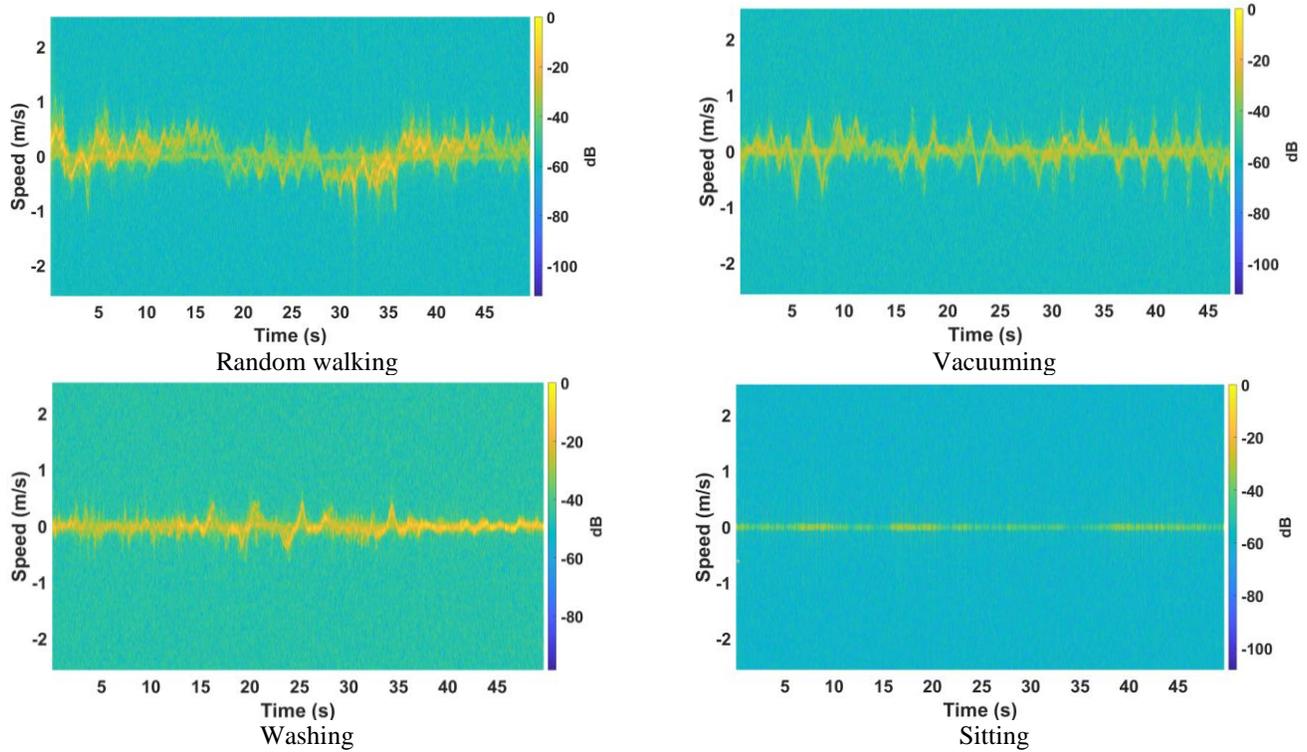

**Fig. 11.** *Time series inputs:* JTF patterns of the subject at his home (a) walking randomly (b) vacuuming (c) washing dishes (d) sitting on the sofa.

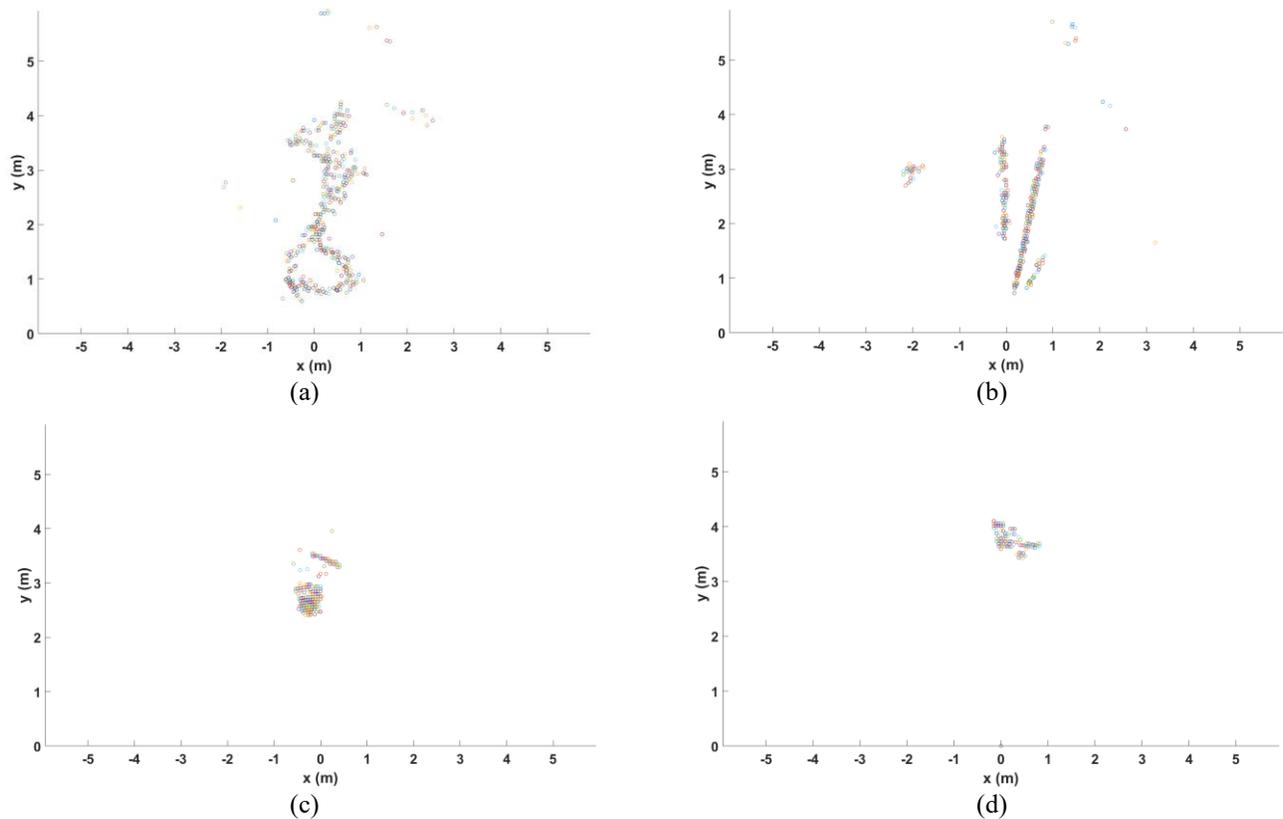

**Fig. 12.** Trajectory of the subject at his home (a) walking randomly (b) vacuuming (c) washing dishes (d) sitting on the sofa.

*1) Deep Learning Results*

The structure of the deep GRU networks, shown in Fig. 4, is applied based on the four defined scenarios. Confusion matrices yielded from datasets collected in the large low clutter environment (scenario # 1 and # 2) are provided in Fig. 13. For both cases (using an independent session and a complete new subject), the network reaches 99% and 98% accuracy, respectively. However, for more complicated scenarios of free-living in-home activities, the system achieves 93% accuracy for scenario #3 when for both training and test sets, samples of both subjects were used (session independent). Also, for a complete new person, the accuracy degrades to 88%. As confusion matrices of scenarios # 3 and #4 in Fig. 14 (a) and (b) indicate, for both scenarios, the network can easily identify an empty room and sedentary behaviour. Moreover, as seen from the precision, F1-score and recall of each class illustrated in Fig. 15, only a few samples were incorrectly identified as walking cycles. This is one of the strengths of the proposed network, as the purpose is to accurately identify walking cycles. In other words, it is crucial to perform a gait extraction algorithm on the actual walking to provide a true prediction/ prevention from gait patterns. Comparing the results of the four defined scenarios, the experiments provide more major insights. Firstly, for the case of data being collected for straight-line walking samples and in a large area without any clutter, the network reaches a high accuracy because of the simplicity of the scenario. Secondly, for a large environment and simple straight line walking datasets, the network performance of the session independent and complete new subject samples shows very small degradation. This is because of the simplicity of samples, whereas this is not true for a real-life in-home application. Thirdly, for in-home free-living activities (for the session independent dataset), if the network is trained with two subjects, it can identify and predict their activities more accurately in future. Lastly, for the more complex and comprehensive scenario, although the accuracy of the system slightly degrades for a complete new subject, it misidentified only a few non-walking samples as walking. Results also show that the F1-score of the classes of vacuuming and washing is more than 80%. Since the input of this system is the time-varying signature of human activities, the proposed system is independent of spatial information. Additionally, although micro-Doppler patterns are dependent on the relative angle to the radar sensor, we showed that the GRU networks could overcome this issue. Regardless of the direction of walking and doing different activities, if the network is trained from the spectrogram of different activities, it then predicts any new scenarios. Therefore, without any restriction, we can use the proposed system for a complete new subject in a new environment without the need for an expensive high-resolution radar.

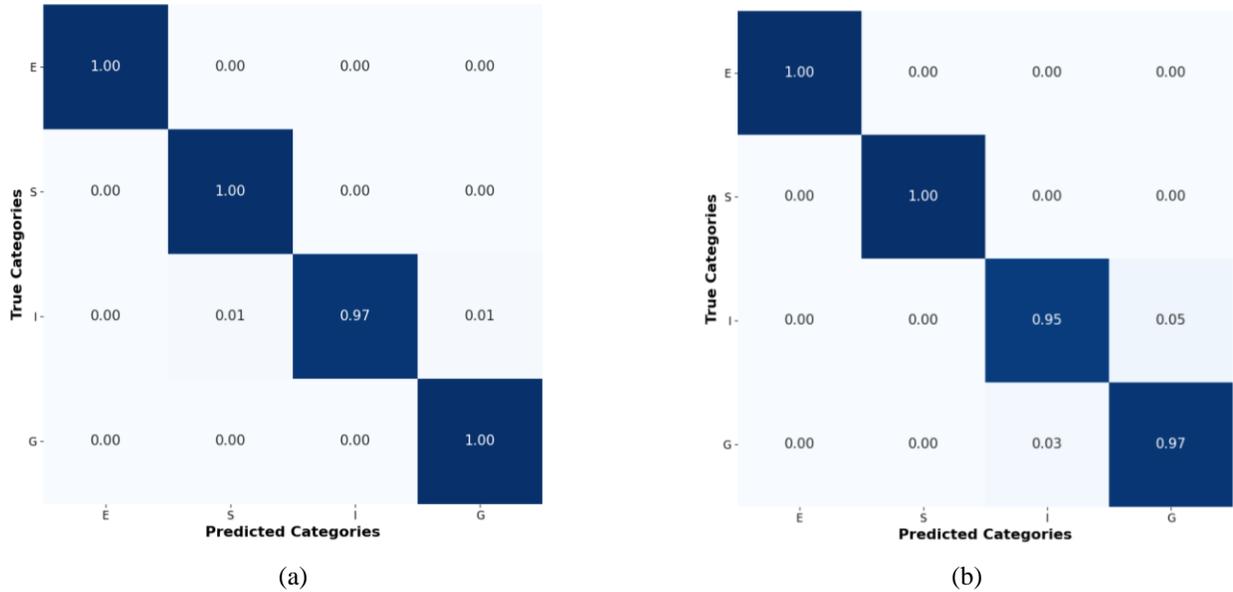

**Fig. 13.** Confusion matrix yielded by the GRU method applied to test datasets in a large low clutter environment (a) trained with datasets collected from 2 subjects and testes based on samples of the two subjects collected at different sessions (b) trained with subject "A" and tested with completely new samples of subject "B". Note that "E", "S", "I", and "G" stand for Empty, Sedentary, In-place movement, and Walking, respectively.

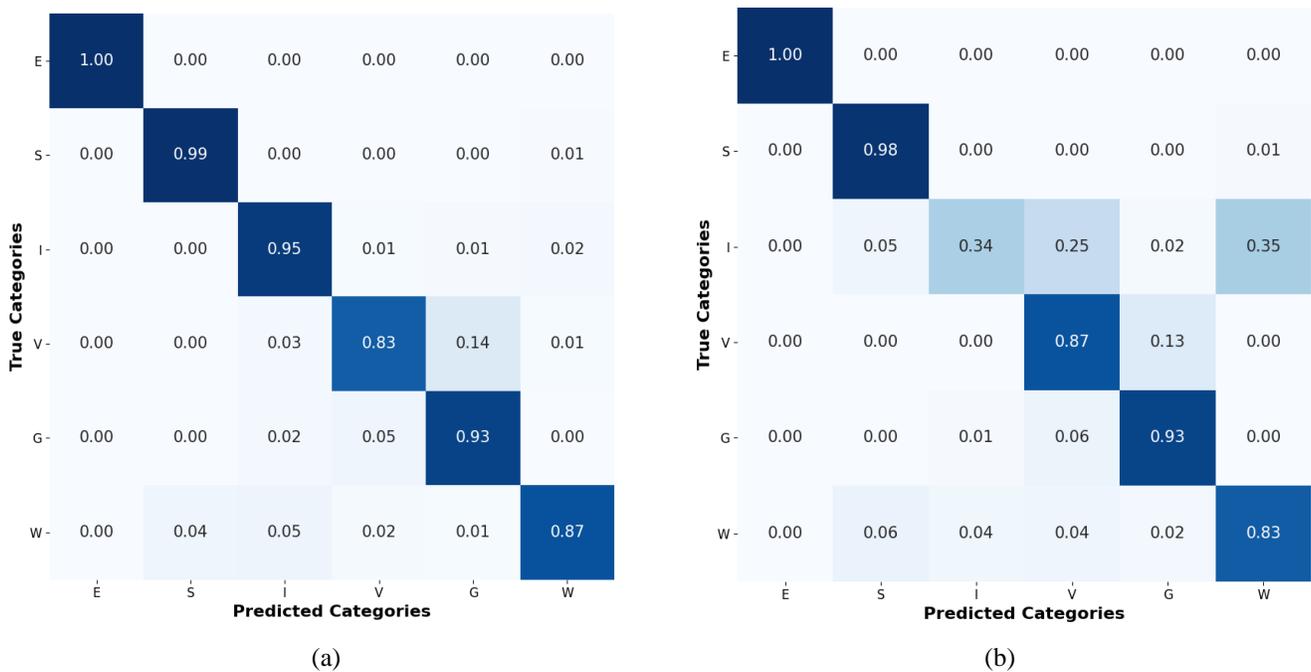

**Fig. 14.** Confusion matrix yielded by the GRU method applied to test datasets collected at the small apartment (a) trained with datasets collected from 2 subjects and tested based on samples of the two subjects collected at a different day (b) trained with subject "A" and tested with a completely new sample of subject "B". Note that "E", "S", "I", "V", "G" and "W" stand for Empty, Sedentary, In-place movement, Vacuuming, Walking and Washing, respectively.

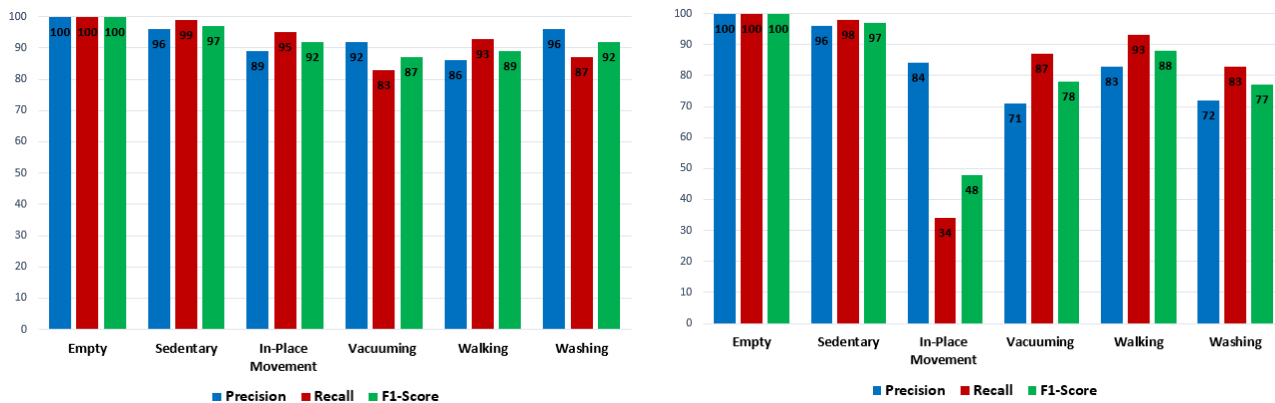

**Fig. 15.** Precision, F1-Score and recall of the GRU applied to test datasets collected at the small apartment (a) trained with datasets collected from 2 subjects and tested based on samples of the two subjects collected at a different day (b) trained with subject "A" and tested with a completely new sample of subject "B".

*2) Output of the Proposed AI-GM&AR System*

As mentioned in Section III. A, the GRU network is trained and optimized in a local machine. The model is then deployed into the cloud to be used in the run-time section.

Preprocessing is performed in the Raspberry Pi to prepare inputs for the PAD algorithm and the GRU network. Then, the real-time preprocessed data is sent to the Azure IoT Hub, using Azure IoT Edge Runtime modules [50]. In Azure, data is collected and stored in the Azure SQL Database [51].

Ultimately, the status of the subject at each room is stored in the storage for the real-time BI (Business Intelligence) dashboard, as shown in Fig. 16. As shown, the BI dashboard consists of four sections: (1) a daily activity report depicted as a bar chart, (2) a gait speed report shown as a line chart, (3) current status illustrated as a predefined icon and (4) extracted gait parameters. Therefore, the proposed cloud-based system provides a report of the subject's daily activity, tracks the current status, captures and records gait parameters over time. Note that the gait extraction algorithms are beyond the scope of this paper. We will cover in-home non-straight-line gait extraction methods in our future publications.

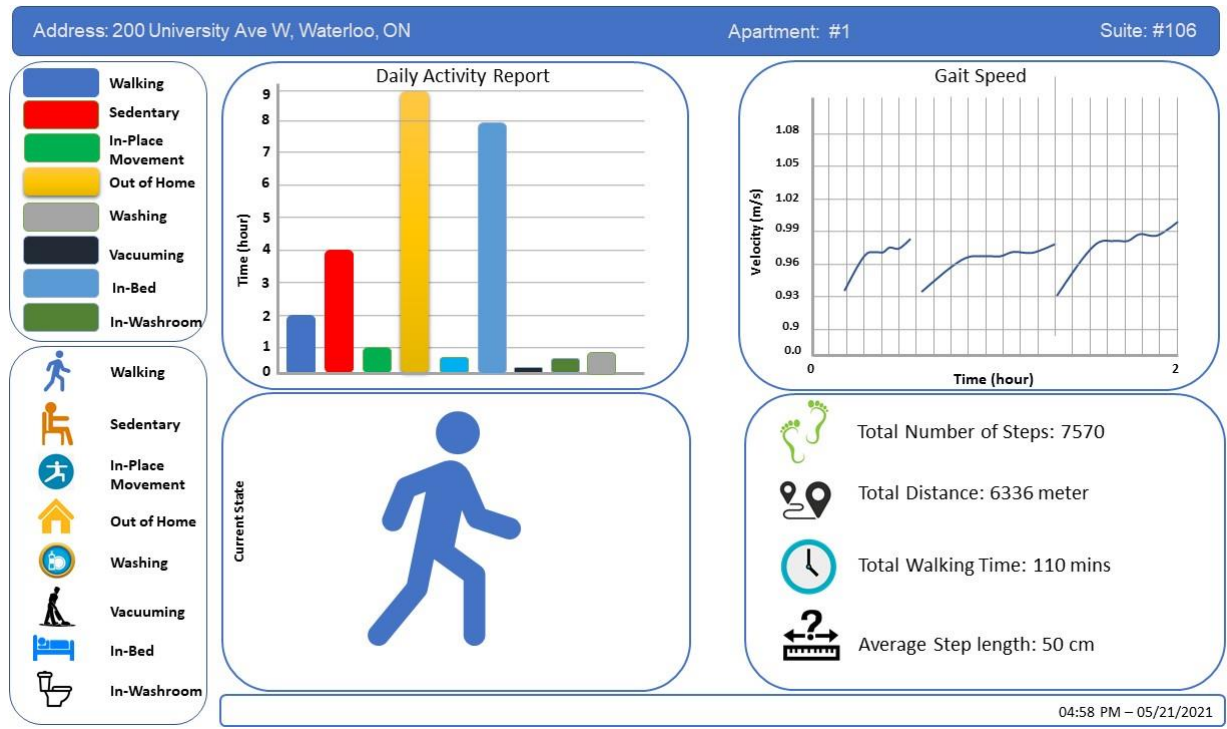

**Fig. 16.** Prototype of the output of our BI dashboard of the proposed AI-GM&AR system. The BI dashboard consists of four sections: a daily activity report, a gait speed, the current status of the subject and extracted gait parameters.

## VI. CONCLUSION

In this paper, we proposed an IoT-based, in-home gait monitoring and activity recognition system. We used mm-Wave FMCW radar sensors coupled with sequential deep learning to generate stream data of human free-living in-home activities. By leveraging the wealth of continuous data, the proposed system can identify the type of activity a person is performing. This system is poised to be a significant achievement in the development of autonomous continuous human monitoring systems as it not only identifies walking cycles and recognizes the type of activity but also has the potential to report the activity level of the subject such as sedentary vs. active, in addition to the washroom frequency and sleep time. We built a first-of-its-kind mm-Wave human in-home free-living activities dataset. Compared to existing datasets for human activity recognition and gait analysis collected at constrained and limited large environments on simple non-real activities and straight-line walking cycles, we monitored in-home free-living human activities. We showed that the joint-time-frequency signature of a single subject performing various activities in his living environment is sufficient and rich information to be delivered to sequential deep learning. We used deep GRU networks to identify in-home free-living activities. We followed two methods of network evaluation: session independent, the network assessment is done with newly collected samples, but the subjects were seen, and assessment based on a complete new subject. We showed that using the common and simple datasets, our deep GRU networks reach 98% accuracy even for a complete new subject. Moreover, for real in-home free-living detests, the proposed network is able to achieve a higher accuracy of 93% for session independent datasets and 88% accuracy even for a completely new subject. We plan to make an in-depth study to improve mm-Wave human activity recognition and gait monitoring under more dynamic scenarios.